\documentstyle[twocolumn,aps,epsf]{revtex}
\begin{document}

\twocolumn[\hsize\textwidth\columnwidth\hsize\csname
@twocolumnfalse\endcsname

\title{Understanding Pentaquark States in QCD}
\author{Shi-Lin Zhu\\
Department of Physics, Peking University, Beijing 100871, China}
\maketitle

\begin{abstract}
We estimate the mass of the pentaquark state with QCD sum rules
and find that pentaquark states with isospin $I=0, 1, 2$ lie close
to each other around $(1.55\pm 0.15)$ GeV. The experimentally
observed baryon resonance $\Theta^+ (1540)$ with $S=+1$ can be
consistently identified as a pentaquark state if its $J^P={1\over
2}^-$. Such a state is expected in QCD. If its parity is positive,
this pentaquark state is really exotic. Now the outstanding issue
is to determine its quantum numbers experimentally.

\end{abstract}
\vspace{0.3cm}
]

\pagenumbering{arabic}

Quantum Chromodynamics (QCD) is believed to be the underlying
theory of the strong interaction. In the high energy regime, QCD
has been tested up to $1\%$ level. In the low energy sector, QCD
is highly nonperturbative due to the non-abelian SU$_c$(3) color
group structure. It is very difficult to calculate the whole
hadron spectrum from first principles in QCD. With the rapid
development of new ideas and computing power, lattice gauge theory
may provide the final solution to the spectrum problem in the
future. But now, people have just been able to understand the
first orbital and radial excitations of the nucleon with lattice
QCD in the baryon sector \cite{liu}.

Under such a circumstance, various models which are QCD-based or
incorporate some important properties of QCD were proposed to
explain the hadron spectrum and other low-energy properties. Among
them, it is fair to say that quark model has been the most
successful one. It is widely used to classify hadrons and
calculate their masses, static properties and low-energy reactions
\cite{isgur}. According to quark model, mesons are composed of a
pair of quark and anti-quark while baryons are composed of three
quarks. Both mesons and baryons are color singlets. Most of the
experimentally observed hadrons can be easily accommodated in the
quark model. Any state with the quark content other than $q\bar q$
or $q q q$ is beyond quark model, which is termed as
non-conventional or exotic. For example, it is hard for
$f_0(980)/a_0(980)$ to find a suitable position in quark model.
Instead it could be a kaon molecule or four quark state
\cite{pdg}.

However, besides conventional mesons and baryons, QCD itself does
not exclude the existence of the non-conventional states such as
glueballs ($gg, ggg, \cdots$), hybrid mesons ($q\bar q g$), and
other multi-quark states ($qq\bar q \bar q$, $qqqq\bar q$,
$qqq\bar q \bar q \bar q$, $qqqqqq, \cdots$). In fact, hybrid
mesons can mix freely with conventional mesons in the large $N_c$
limit \cite{cohen}. In the early days of QCD, Jaffe proposed the H
particle \cite{jaffe} with MIT bag model, which was a six quark
state. Unfortunately it was not found experimentally.

In the past years there have accumulated some experimental
evidence of possible existence of glueballs  and hybrid mesons
with exotic quantum numbers like $J^{PC}=1^{-+}$ \cite{pdg}.
Recently BES collaboration observed a possible signal of a proton
anti-proton baryonium in the $J/\Psi$ radiative decays \cite{bes}.
But none of these states has been pinned down without controversy.

Recently LEPS Collaboration at the SPring-8 facility in Japan
observed a sharp resonance $\Theta^+$ at $1.54\pm 0.01$ GeV with a
width smaller than 25 MeV and a statistical significance of
$4.6\sigma$ in the reaction $\gamma n \to K^+ K^- n$ \cite{leps}.
This resonance decays into $K^+ n$, hence carries strangeness
$S=+1$. Later DIANA Collaboration at ITEP observed the same
resonance at $1539\pm 2$ MeV with a width less than $9$ MeV in a
different reaction $K^+ Xe\to \Theta^+ Xe^\prime \to K^0 p
Xe^\prime$ \cite{diana}. Now $\Theta^+$ decays into $K^0 p$. The
convincing level is $4.4\sigma$. Very recently CLAS Collaboration
in Hall B at JLAB also observed $\Theta^+$ in the $K^+ n$
invariant mass at $1542\pm 5$ MeV in the exclusive measurement of
the $\gamma d \to K^+ K^- pn $ reaction \cite{clas}. The
statistical significance is $5.3\sigma$. The measured width is
$21$ MeV, consistent with CLAS detector resolution. The isospin,
parity and angular moment of the $\Theta^+$ particle has not been
determined rigorously yet. There is only preliminary hint that
$\Theta^+$ might be an iso-singlet from the featureless $M(K^+ p)$
spectrum in the CLAS measurement.

Diakonov et al. proposed the possible existence of the $S=1$
$J^P={1\over 2}^+$ resonance at $1530$ MeV with a width less than
15 MeV using the chiral soliton model \cite{diak}, which partly
motivated the recent experimental search of this particle. The
authors argued that $\Theta^+$ is the lightest member of the
anti-decuplet multiplet which is the third rotational state of the
chiral soliton model. Assuming that the $N(1710)$ is a member of
the anti-decuplet, $\Theta^+$ mass is fixed with the symmetry
consideration of the model. Gao and Ma also discussed the
experimental search of this pentaquark state \cite{gao}. Recently
Polyakov and Rathke suggested possible photo-excitation of the
pentaquark states \cite{poly}.

Capstick, Page and Winston pointed out that identifying $N(1710)$
as a member of the anti-decuplet in the chiral soliton model is
kind of arbitrary \cite{page}. Instead, if the anti-decuplet
$P_{11}$ is $N(1440)$, the $\Theta^+$ would be stable as the
ground state octet with a very low mass while $\Theta^+$ would be
very broad with the anti-decuplet $P_{11}$ being $N(2100)$
\cite{page}. Furthermore, if the decay width of the anti-decuplet
$N(1710)$ was shifted upwards to be comparable with PDG values,
the predicted width of $\Theta^+$ particle would have exceeded the
present experimental upper bound \cite{page}. The authors
hypothesized that $\Theta^+$ could be an isotensor resonance with
negative parity \cite{page}. The isospin violating decay width is
naturally very small. Hence $\Theta^+$ is a narrow state. Note
similar isospin violating decay mechanism makes $D_{sJ}(2317)$ and
$D_{sJ}(2460)$ states very narrow \cite{babar,cleo,zhu}.

Stancu and Riska studied the stability of the strange pentaquark
state assuming a flavor-spin hyperfine interaction between quarks
in the constituent quark model \cite{riska}. They suggested that
the lowest lying p-shell pentaquark state with positive parity
could be stable against strong decays if the spin-spin interaction
betwenn the strange antiquark and up/down quark was strong enough
\cite{riska}. The group classification of strange pentaquarks is
given in \cite{group}.

In Ref. \cite{hosaka} Hosaka emphasized the important role of the
hedgehog pion in the strong interaction dynamics which drives the
formation of the $\Theta^+$ particle and the detailed energy level
ordering. Hyodo, Hosaka and Oset suggested measuring the
$\Theta^+$ quantum numbers using the reaction $K^+ p\to \pi^+ K^+
n$ \cite{oset}.

In Ref. \cite{lipkin} Karliner and Lipkin discussed the dynamics
of the diquark-triquark pentaquark state with a rough estimate of
the mass of $1592$ MeV and $I=0, J^P={1\over 2}^+$ in the
constituent quark model. The same authors discussed pentaquarks
containing a heavy baryon \cite{lip}.

Very recently Jaffe and Wilczek suggested that the observed
$\Theta^+$ state could be composed of an anti-strange quark and
two highly correlated up and down quark pairs arising from strong
color-spin correlation force \cite{bob}. The resulting $J^P$ of
$\Theta^+$ is ${1\over 2}^+$.

Due to the low mass of $\Theta^+$, we think its angular momentum
is likely to be one half. In this paper we shall employ QCD sum
rules to estimate the mass of the pentaquark states with
$J={1\over 2}$ and $I=0, 1, 2$.

The method of QCD sum rules incorporates two basic properties of
QCD in the low energy domain: confinement and approximate chiral
symmetry and its spontaneous breaking. One considers a correlation
function of some specific interpolating currents with the proper
quantum numbers and calculates the correlator perturbatively
starting from high energy region. Then the resonance region is
approached where non-perturbative corrections in terms of various
condensates gradually become important. Using the operator product
expansion, the spectral density of the correlator at the quark
gluon level can be obtained in QCD. On the other hand, the
spectral density can be expressed in term of physical observables
like masses, decay constants, coupling constants etc at the hadron
level. With the assumption of quark hadron duality these two
spectral densities can be related to each other. In this way one
can extract hadron masses etc. For the past decades QCD sum rules
has proven to be a very powerful and successful non-perturbative
method \cite{svz,reinders}.

We use the following interpolating current for the $I=2$
pentaquark state where three quarks and the remaining $q\bar q$
pair are both in a color adjoint representation. We note in
passing such a choice of color configuration is not unique
\cite{color}.
\begin{equation}\label{cu1}
\eta_2(x)=\epsilon^{abc} [u^T_a(x) C\gamma_\mu u_b (x)] \gamma^\mu
\gamma_5 u_e (x) {\bar s}_e (x) i\gamma_5 u_c(x)
\end{equation}
for a charge $Q=+3$ state. Or
\begin{eqnarray}\label{cu2} \nonumber
\eta_2(x)={1\over \sqrt{2}}\{ \epsilon^{abc} [u^T_a(x) C\gamma_\mu
u_b (x)] \gamma^\mu \gamma_5 d_e (x)\\
\times {\bar s}_e (x) i\gamma_5 d_c(x) +\left( u\leftrightarrow
d\right) \}
\end{eqnarray}
for a charge $Q=+1$ state. The isospin of the current is shown in
its lower index. $a, b, c$ etc are the color indices. $T$ denote
transpose. $C$ is the charge conjugation matrix. The property
$\left( C\gamma_\mu\right)^T =C\gamma_\mu$ ensures the current is
symmetric under the exchange of the two quark fields inside the
bracket. Both currents in Eq. (\ref{cu1}) and (\ref{cu2}) carry an
isospin of two.

We want to emphasize the isospin and color structure of these
currents guarantee they will never couple to a $K^+ n$ molecule or
any other $K^+ n$ intermediate states since the isospin of a $K^+
n$ system can only be zero or one. One may worry about $K \Delta$
intermediate states since their total isospin can be two. However
the energy of $K \Delta$ intermediate states are greater than $m_K
+ m_\Delta =1726$, which is much larger than observed $\Theta^+$
mass. In some cases, if the energy of the S-wave intermediate
state is significantly lower than the resonance, its contribution
could turn out to be quite important \cite{zhu1,zhu}. Angular
momentum and parity conservation requires that only D-wave $K
\Delta$ intermediate states could contribute if $J^P$ of
$\Theta^+$ is ${1\over 2}^-$. When its $J^P$ is ${1\over 2}^+$,
then the $K\Delta$ combination should be of p-wave. Either P-wave
or D-wave $K \Delta$ continuum contribution is negligible compared
with the lower $\Theta^+$ narrow resonance contribution. So these
currents couple to $I=2$ strange pentaquark states.

For $I=0$ pentaquark state the interpolating current is
\begin{eqnarray}\label{cu3} \nonumber
\eta_0(x)={1\over \sqrt{2}} \epsilon^{abc} [u^T_a(x) C\gamma_5
d_b (x)] \{ u_e (x)\\
\times {\bar s}_e (x) i\gamma_5 d_c(x) - \left( u\leftrightarrow
d\right) \}
\end{eqnarray}
where $\left( C\gamma_5\right)^T =-C\gamma_5$ ensures the isospin
of the up and down quark pair inside the first bracket to be zero.
The anti-symmetrization in the second bracket ensures that the
isospin of the other up and down quark pair is also zero.

For $I=1$ pentaquark state the interpolating currents are
\begin{eqnarray}\label{cu4} \nonumber
\eta_1(x)={1\over \sqrt{2}} \epsilon^{abc} [u^T_a(x) C\gamma_\mu
d_b (x)] \{ \gamma^\mu \gamma_5 u_e (x)\\
\times {\bar s}_e (x) i\gamma_5 d_c(x) -\left( u\leftrightarrow
d\right) \}
\end{eqnarray}
or
\begin{eqnarray}\label{cu5} \nonumber
\eta_1(x)={1\over \sqrt{2}} \epsilon^{abc} [u^T_a(x) C\gamma_5
d_b (x)]  \{ u_e (x)\\
\times {\bar s}_e (x) i\gamma_5 d_c(x) +\left( u\leftrightarrow
d\right) \}
\end{eqnarray}
In the following we use currents (\ref{cu2}), (\ref{cu3}) and
(\ref{cu5}) to perform the calculation.

We introduce the overlapping amplitude $f_j$ of the interpolating
currents with the corresponding pentaquark states.
\begin{eqnarray}
\label{decay} \langle 0|\eta_j(0)|p, I=i\rangle= f_j u(p)
\end{eqnarray}
where $u(p)$ is the Dirac spinor of pentaquark field with $I=j$.

We consider the correlator
\begin{equation}\label{cor-1}
i\int d^4 x e^{ipx} \langle 0|T\left (\eta_j (x), {\bar \eta}_j
(0) \right )|0\rangle\ = \Pi (p) {\hat p} + \Pi' (p )
\end{equation}
where $\bar \eta =\eta^\dag \gamma_0$ and ${\hat p} =p_\mu \cdot
\gamma^\mu$. Throughout this note we focus on the chirality even
structure $\Pi (p)$. At the hadron level it can be written as
\begin{equation}
\Pi (p)= {f^2_j \over p^2 -M_j^2} + \mbox{higher states}
\end{equation}
where $M_j$ is the pentaquark mass.

On the other hand, it will be calculated in terms of quarks and
gluons. For example, for the $I=2$ case, the correlator reads
\begin{eqnarray}\label{x}\nonumber
i\int e^{ipx} dx \{-2\epsilon^{abc}\epsilon^{a'b'c'} {\bf
\mbox{Tr}}\left[ iS_{aa'}(x)\cdot \gamma_\nu C \cdot iS^T_{bb'}(x)
 \cdot C\gamma_\mu \right]\\  \nonumber \times {\bf \mbox{Tr}}\left[
i\gamma_5  \cdot iS^s_{e'e}(-x) \cdot i\gamma_5  \cdot
iS_{cc'}(x)\right] \gamma^\mu \gamma_5  \cdot iS_{ee'}(x) \cdot
\gamma^\nu \gamma_5 \\ \nonumber +2\epsilon^{abc}\epsilon^{a'b'c'}
{\bf \mbox{Tr}}\left[
iS_{aa'}(x) \cdot \gamma_\nu C  \cdot iS^T_{bb'}(x) \cdot  C\gamma_\mu \right]\\
\nonumber  \times \gamma^\mu \gamma_5  \cdot iS_{ec'}(x)  \cdot
i\gamma_5 \cdot iS^s_{e'e}(-x) \cdot i\gamma_5  \cdot iS_{ce'}(x)
\cdot \gamma^\nu \gamma_5 \\ \nonumber
+4\epsilon^{abc}\epsilon^{a'b'c'}\gamma^\mu \gamma_5
 \cdot iS_{ea'}(x) \cdot  \gamma_\nu C  [ iS_{bc'}(x) \cdot i\gamma_5
 \times \\  \nonumber iS^s_{e'e}(-x) \cdot i\gamma_5 \cdot  iS_{cb'}(x)]^T C\gamma_\mu
 \cdot iS_{ae'}(x)  \cdot \gamma^\nu\gamma_5 \}
\end{eqnarray}
where $iS^s_{e'e}(-x)$ is the strange quark propagator in the
coordinate space.

After making Fourier transformation to the above equation and
invoking Borel transformation to Eq. (\ref{cor-1}) we get
\begin{equation}\label{b}
f^2_je^{-{M_j^2\over T^2}}=\int_{m_s^2}^{s_0} e^{-{s\over T^2}}
\rho_j (s)
\end{equation}
where $m_s$ is the strange quark
mass, $\rho_j (s)$ is the spectral density and $s_0$ is the
threshold parameter used to subtract the higher state contribution
with the help of quark-hadron duality assumption. Roughly speaking
$\sqrt{s_0}$ is around the first radial excitation mass. The
spectral density reads
\begin{eqnarray}\label{spectral}\nonumber
\rho_2={s^5\over  4^6 7! \pi^8}+{s^3\over  2^{13} 6! \pi^8}
\langle g_s^2 G^2 \rangle +{s^2\over 1536 \pi^4} [ {1\over 3}
\langle \bar q q\rangle^2 \\ \nonumber +{7\over 6}\langle \bar
qq\rangle \langle \bar ss\rangle ]+{5\over 108}\langle \bar q
q\rangle^3 \langle \bar s s\rangle \delta (s)  \\ \nonumber
\rho_1={5 s^5\over  2^{17} 7! \pi^8}+{s^2\over 1536 \pi^4} [
{13\over 48} \langle \bar q q\rangle^2  +{1\over 3}\langle \bar
qq\rangle\langle \bar ss\rangle ]\\ \nonumber +[{1\over
108}\langle \bar q q\rangle^3 \langle \bar s s\rangle -{1\over
576}\langle \bar q q\rangle^4] \delta (s)  \\ \nonumber \rho_0={3
s^5\over  4^8 7! \pi^8}+{s^2\over 1536 \pi^4} [ {5\over 12}
\langle \bar q q\rangle^2  +{11\over 24}\langle \bar
qq\rangle\langle \bar ss\rangle ]\\ \nonumber +[{7\over
432}\langle \bar q q\rangle^3 \langle \bar s s\rangle +{1\over
864}\langle \bar q q\rangle^4] \delta (s)
\end{eqnarray}
where we have used the factorization approximation for the
multi-quark condensates.

In order to extract $M_j$ we first take derivative of Eq.
(\ref{b}) with respect to $1/T^2$. Then we divide it by Eq.
(\ref{b}) to get
\begin{eqnarray}
M_j^2 =\frac{\int_{m_s^2}^{s_0} d s e^{-{s/T^2}} \rho'
(s)}{\int_{m_s^2}^{s_0} d s e^{-{s/T^2}} \rho (s)}
\end{eqnarray}
with $\rho'(s)=s\rho(s)$ except that $\rho'(s)$ does not contain
the last term in $\rho(s)$.

In the numerical analysis of the sum rules the values of various
QCD condensates are $\langle\bar ss\rangle=-(0.8\pm 0.1)*(0.24
~\mbox{GeV})^3$, $ \langle g_s^2 GG\rangle=0.48 ~\mbox{GeV}^4$. We
use $m_s (1 \mbox{GeV}) =0.15$ GeV for the strange quark mass in
the ${\bar {MS}}$ scheme.

\begin{figure}
\epsfxsize=8.5cm \centerline{\epsffile{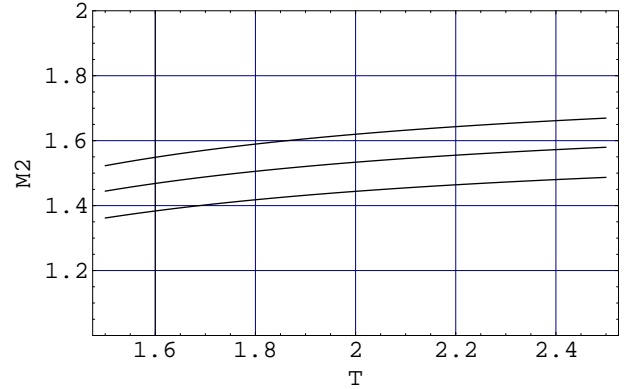}} \vspace{0.5cm}
\caption{The variation of the $I=2$ pentaquark state mass $M_2$
with the Borel parameter $T$ (in unit of GeV) and the continuum
threshold $s_0$. From bottom to top the curves correspond to
$s_0=3.61, 4.0, 4.41$ GeV$^2$ respectively. } \label{fig1}
\end{figure}

\begin{figure}
\epsfxsize=8.5cm \centerline{\epsffile{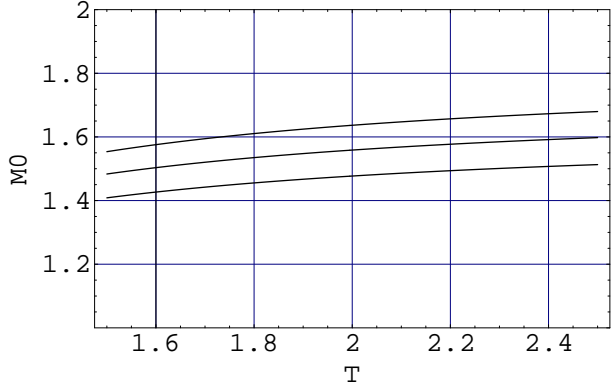}}
\vspace{0.5cm}\caption{The variation of the $I=0$ pentaquark state
mass $M_0$ with the Borel parameter $T$ (in unit of GeV) and the
continuum threshold $s_0$. From bottom to top the curves
correspond to $s_0=3.61, 4.0, 4.41$ GeV$^2$ respectively. }
\label{fig3}
\end{figure}

Numerically we have $M_2 =(1.53\pm 0.15) \mbox{GeV}$, $M_1
=(1.59\pm 0.15)\mbox{GeV}$, $ M_0 =(1.56\pm 0.15) \mbox{GeV}$,
where the central value corresponds to $T=2$ GeV and $s_0=4$
GeV$^2$. The variation of $M_2, M_0$ with both $T$ and $s_0$ is
shown in Figures (\ref{fig1})-(\ref{fig3}), which contributes to
the errors of the extracted value, together with the truncation of
the operator product expansion, the uncertainty of vacuum
condensate values and the factorization approximation of the
multiquark condensates. In the working interval of the Borel
parameter $M_j$ is reasonably stable with $T$. However the level
ordering of the qentaquarks with isospin should not be taken too
seriously due to the large uncertainty.

Our results suggest that strange pentaquark states with $I=0, 1,
2$ lie close to each other, which is very different from the
prediction of chiral soliton model where only the lightest $I=0$
member is around $1540$ MeV and all other states are several
hundred MeV higher \cite{diak}. As we mentioned before, our
interpolating currents can couple to pentaquark states with either
positive or negative parity. QCD sum rules approach only picks out
the lowest mass pentaquark state in the specific channel. It is
impossible to determine its parity in our formalism.

One may tend to think that $\Theta$ is a $J^P={1\over 2}^-$
strange pentaquark state from previous quark model experience.
One radial or orbital excitation of constituent quarks typically
carries energy around $400-500$ MeV. Naively one would expect the
strange pentaquark mass with positive parity to be around $4M_u +
M_s + 400\sim 2050$ MeV, where $M_u=300$ MeV and $M_s=450$ MeV are
roughly the up and strange quark constituent mass. The additional
$\sim 400$ MeV comes from one orbital excitation for the
positive-parity $\Theta^+$. If future experiments further
establish the $\Theta^+$ parity to be positive, then this particle
is really very exotic. Otherwise, a $J^P={1\over 2}^-$ $\Theta^+$
particle is an expected strange pentaquark state in QCD. The
outstanding issue is to measure the quantum numbers of the
$\Theta^+$ particle, especially the parity, angular momentum and
isospin in the future experiments.

As a byproduct, we want to mention that the radial excitation
$\Theta'$ of the strange pentaquark may lie around $2.0$ GeV. If
$\Theta$ is an iso-tensor state as suggested in \cite{page},
$\Theta'$ should mainly decay into final states containing one or
two pions like $\Theta'\to K^+\Delta\to \pi^0 K^+ n, \Theta'\to
\Theta \pi\pi\to K^+ n \pi\pi$, which conserve isospin symmetry.
The isospin conserving decay mode $\Theta\to \pi^0 K^+ n$ is
kinematically forbidden for an isotensor pentaquark. If its
isospin is $0$ or $1$, then the dominant mode should be
$\Theta'\to K^+ n, \Theta'\to \Theta \pi\pi\to K^+ n \pi\pi$. In
other words, the decay mode $\Theta'\to \pi^0 K^+ n$ with a single
pion in the final state can be used to distinguish the isospin of
the strange pentaquark if enough events are collected in the
future experiments.

This project was supported by the National Natural Science
Foundation of China, Ministry of Education of China, FANEDD and
SRF for ROCS, SEM.



\begin{thebibliography}{99}
\bibitem{liu}S. J. Dong et al., hep-ph/0306199.
\bibitem{isgur}S. Godfrey and N. Isgur, Phys. Rev. D 32, 189
(1985);S. Capstick and N. Isgur, Phys. Rev. D 34, 2809 (1986).
\bibitem{pdg}Particle Dada Group, Phys. Rev. D 66, 010001 (2002).
\bibitem{cohen}T. D. Cohen, Phys. Lett. B 427, 348 (1998).
\bibitem{jaffe}R. L. Jaffe, Phys. Rev. Lett. 38, 195 (1977).
\bibitem{bes}J. Z. Bai et al, BES Collaboration, Phys. Rev. Lett.
91, 022001 (2003).
\bibitem{leps}T. Nakano et al., Phys. Rev. Lett. 91, 012002
(2003).
\bibitem{diana}V. V. Barmin et al., hep-ex/0304040.
\bibitem{clas}S. Stepanyan et al., hep-ex/0307018.
\bibitem{diak}D. Diakonov, V. Petrov, and M. Ployakov, Z. Phys. A
359, 305 (1997).
\bibitem{gao}H. Gao and B.-Q. Ma, hep-ph/0305294, Mod. Phys. Lett.
A 14, 2313 (1999).
\bibitem{poly}M. V. Ployakov and A, Rathke, hep-ph/0303138.
\bibitem{page}S. Capstick, P. R. Page, and W. Roberts,
hep-ph/0307019.
\bibitem{babar}B. Aubert et al., Babar Collaboration,
hep-ex/0304021, Phys. Rev. Lett. 90, 242001 (2003).
\bibitem{cleo}D. Besson et al., hep-ex/0305017.
\bibitem{zhu}Y. B. Dai, C. S. Huang, C. Liu and S.-L. Zhu,
hep-ph/0306274, Phys. Rev. D (in press).
\bibitem{riska}F. Stancu and D. O. Riska, hep-ph/0307010.
\bibitem{group}B. G. Wybourne, hep-ph/0307170.
\bibitem{hosaka}A. Hosaka, hep-ph/0307232.
\bibitem{oset}T. Hyodo, A. Hosaka, E. Oset, nucl-th/0307105.
\bibitem{lipkin}M. Karliner and H. J. Lipkin, hep-ph/0307243.
\bibitem{lip}M. Karliner and H. J. Lipkin, hep-ph/0307343.
\bibitem{bob}R. Jaffe and F. Wilczek, hep-ph/0307341.
\bibitem{svz}M.A. Shifman, A.I. Vainshtein and V.I. Zakharov, Nucl. Phys. {\bf
B 147}, 385 (1979).
\bibitem{reinders}L. J. Reinders, H. Rubinstein, and S. Yazaki,
Phys. Rept. 127, 1 (1985).
\bibitem{color}H. Hogaasen and P. Sorba, Nucl. Phys. B 145, 119
(1978).
\bibitem{zhu1}Shi-Lin Zhu and Yuan-Ben Dai, Mod. Phys. Lett. A 14, 2367 (1999).
\end{thebibliography}
\end{document}